\documentclass[aps,prl,twocolumn,showpacs,superscriptaddress]{revtex4}
\usepackage{bm,amsmath,amssymb,latexsym,mathrsfs,enumerate,color}
\usepackage[mathcal]{euscript}
\usepackage[hypertex]{hyperref}
\usepackage{graphicx}
\graphicspath{{figs/}}
\renewcommand{\vec}[1]{\bm{#1}}
\preprint{\textcolor{blue}{\texttt{version No $3.141592$}}}

\begin{document}

\title{Switching phenomena in magnetic vortex dynamics}

\author{Yuri B. Gaididei}
 \affiliation{Institute for Theoretical Physics, 03680 Kiev, Ukraine}
 \author{Volodymyr P. Kravchuk}
 \affiliation{National Taras Shevchenko University of Kiev, 03127 Kiev, Ukraine}
\author{Franz G.~Mertens}
 \affiliation{Physics Institute, University of Bayreuth, 95440 Bayreuth, Germany}
\author{Denis D. Sheka}
 \email[Corresponding author. Electronic address:\\]{denis\_sheka@univ.kiev.ua}
 \affiliation{National Taras Shevchenko University of Kiev, 03127 Kiev, Ukraine}

\date{\today}

%
%

\begin{abstract}
A magnetic nanoparticle in a vortex state is a promising candidate for the information storage. One bit of information corresponds to the upward or downward magnetization of the vortex core (vortex polarity). Generic properties of the vortex polarity switching are insensitive of the way how the vortex dynamics was excited: by an AC magnetic field, or by an electrical current. We study theoretically the switching process and describe in detail its mechanism, which involves the creation and annihilation of an intermediate vortex-antivortex pair.
\end{abstract}




\maketitle

\section*{Introduction}
\label{sec:intro} %

Artificial mesoscopic magnetic structures provide now a wide testing area for concepts of nanomagnetism and numerous prospective applications \cite{Hubert98,Skomski03}. Investigations of magnetic nanostructures include studies of magnetic nanodots, which are submicron disk-shaped particles, which have a single vortex in the ground state due to the competition between exchange and magnetic dipole-dipole interaction \cite{Hubert98}. A vortex state is obtained in nanodots that are larger than a single domain whose size is a few nanometers: e.g. for the Permalloy (Ni$_{80}$Fe$_{20}$) nanodot the exchange length $\ell\approx5$\,nm. Having nontrivial topological structure on the scale of a nanomagnet, the magnetic vortex is a promising candidate for the high density magnetic storage and high speed magnetic random access memory \cite{Cowburn02}. For this one needs to control magnetization reversal, a process in which vortices play a big role \cite{Guslienko01}. Great progress has been made recently with the possibility to observe high frequency dynamical properties of the vortex state in magnetic dots by Brillouin light scattering of spin waves \cite{Demokritov01,Hillebrands02}, time-resolved Kerr microscopy \cite{Park03}, phase sensitive Fourier transformation technique \cite{Buess04}, X-ray imaging technique \cite{Choe04}, and micro-SQUID technique \cite{Thirion03}.

The control of magnetic nonlinear structures using an electrical current is of special interest for applications in spintronics \cite{Tserkovnyak05,Bader06}. The spin torque effect, which is the change of magnetization due to the interaction with an electrical current, was predicted by \citet{Slonczewski96} and \citet{Berger96} in \citeyear{Slonczewski96}. During the last decade this effect was tested in different magnetic systems \cite{Tsoi98,Myers99,Krivorotov05} and nowadays it plays an important role in spintronics \cite{Tserkovnyak05,Marrows05}. Recently the spin torque effect was observed in vortex state nanoparticles. In particular, circular vortex motion can be excited by an AC \cite{Kasai06} or a DC \cite{Pribiag07} spin-polarized current. Very recently it was predicted theoretically \cite{Caputo07,Sheka07b} and observed experimentally \cite{Yamada07} that the vortex polarity can be controlled using a spin-polarized current. This opens up the possibility of realizing electrically controlled magnetic devices, changing the direction of modern spintronics \cite{Cowburn07}.

The basics of the \emph{magnetic vortex} \cite{Kovalev-vortex} statics and dynamics in Heisenberg magnets was studied in 1980s, in particular by the group of Kosevich \cite{Kosevich83,Kosevich85,Kosevich90}. Typically, the vortex is considered as a rigid particle without internal degrees of freedom; such a vortex undergoes a so-called gyroscopic dynamics in the framework of Thiele-like equations \cite{Thiele73,Huber82,Nikiforov83}. Using a rigid vortex dynamics a number of dynamical effects were studied, for a review see Ref.~\cite{Mertens00}. Cycloidal oscillations around the mean vortex trajectory can be explained by taking into account changes of the vortex shape due to its velocity \cite{Mertens00,Kovalev03a}.

The rigid approach also fails when considering the vortex dynamics under the influence of a strong or fast external force. In particular, it is known that external pumping excites internal modes in vortex dynamics in Heisenberg magnets, whose role is important for understanding the switching phenomena \cite{Gaididei99,Gaididei00,Kovalev02,Kovalev03,Zagorodny03}, and the limit cycles in vortex dynamics \cite{Zagorodny04,Sheka05}.

Very recently we have found that under the action of a magnetic field, or, alternatively, under the action of a spin current on a vortex state nanodot, the vortex core magnetization (polarity) can be switched. This effect was studied in Ref.~\cite{Caputo07} analytically and confirmed numerically by direct spin-lattice simulations for Heisenberg magnets. However, the presence of the long-range dipolar interaction crucially changes the vortex dynamics \cite{Sheka07b}, because  this interaction creates an effective nonhomogeneous anisotropy \cite{Caputo07}.  The goal of this paper is to consider the generic properties of polarity switching in magnetic nanodots.

\section{Model and equations of motion}
\label{sec:model} %

The magnetic energy of nanodots consists of two parts: Heisenberg exchange and
dipolar interactions

\begin{equation} \label{eq:Hamiltonian}
\begin{split}
\mathcal{H} &= -\frac{\ell^2}{2}\!\! \sum_{\left(\vec{n},\vec{\delta}\right)}\!
\vec{S}_{\vec{n}}\cdot \vec{S}_{\vec{n}+\vec{\delta}}\\
&+ \frac{1}{8\pi}\!\! \sum_{\substack{\vec{n}, \vec{n}'\\\vec{n} \neq \vec{n}'}}\!
\frac{\vec{S}_{\vec{n}}\cdot \vec{S}_{\vec{n}'}-3
\left(\vec{S}_{\vec{n}}\cdot \vec{e}_{\vec{n}\vec{n}'} \right)
\left(\vec{S}_{\vec{n}'}\cdot \vec{e}_{\vec{n}\vec{n}'} \right)}{
|\vec{n}- \vec{n}'|^3}
\end{split}
\end{equation}
Here $\vec{S}_{\vec{n}}$ is a unit vector which determines the spin direction
at the lattice point $\vec{n}$, $\ell = \sqrt{A/(\mu_0 M_S^2)}$ is the
exchange length ($A$ is the exchange constant, $\mu_0$ is the vacuum
permeability, $M_S$ is the saturation magnetization), the vector
$\vec{\delta}$ connects nearest neighbors, and $\vec{e}_{\vec{n}\vec{n}'}
\equiv (\vec{n} - \vec{n}')/|\vec{n} - \vec{n}'|$ is a unit vector. The
lattice constant is chosen as a unity length.

It is known that as a result of the competition between the exchange and the dipolar  interactions the ground state of a thin magnetically soft nanodisk is a
vortex state: the magnetization lies in the XY disk plane in the main part of
the disk and is parallel to the disk edge, forming the magnetic flux-closure
pattern characterized by the vorticity $q=+1$. At the disk center the
magnetization distribution forms a vortex core, which is oriented either
parallel or antiparallel to the z-axis. The former is characterized by a
polarity $p=+1$ and the latter by $p=-1$.

The spin dynamics of the system is described by the modified
Landau--Lifshitz--Gilbert equation
\begin{equation} \label{eq:LLS-discrete}
\dot{\vec{S}_{\vec{n}}}
= -\vec{S}_{\vec{n}}\times \frac{\partial
\mathcal{H}}{\partial \vec{S}_{\vec{n}}} - \alpha \vec{S}_{\vec{n}}
\times \dot{\vec{S}_{\vec{n}}}
+ \vec{S}_{\vec{n}}\times \vec{b}_{\vec{n}}.
\end{equation}
Here the overdot indicates the derivative with respect to the dimensionless
time $\tau=\omega_0 t$ with $\omega_0 = 4\pi\gamma M_S$, $\alpha\ll1$ is a
damping coefficient. The last term is a spin torque due to external forces,
which results in a magnetic field $\vec{b}_{\vec{n}}$. In the presence of the
homogeneous rotating field $\vec{B}(t) = \left(B\cos\omega \tau,
B\sin\omega\tau, 0\right)$, the dimensionless field
\begin{subequations} \label{eq:b}
\begin{equation} \label{eq:b-field}
\vec{b}=b_x+{i} b_y = \frac{B}{4\pi M_S}\exp({i} \omega\tau).
\end{equation}
Here and below all frequencies are measured in units of $\omega_0$, and all distances in units of $\ell$.

When an electrical current is injected in the pillar structure,
perpendicular to the nanodisk plane, it influences locally the spin
$\vec{S}_{\vec{n}}$ of the lattice through the spin torque
$\vec{T}_{\vec{n}} = \vec{S}_{\vec{n}}\times \vec{b}_{\vec{n}}$
\cite{Slonczewski96,Berger96}, where
\begin{equation} \label{eq:T}
\vec{b}_{\vec{n}}= j\,\sigma\mathcal{A} \frac{\vec{S}_{\vec{n}} \times
\hat{\vec{z}}}{1+\sigma
\mathcal{B} \vec{S}_n\cdot \hat{\vec{z}}}.
\end{equation}
\end{subequations}
Here $j=J_e/J_p$ is a normalized spin current, $J_e$ is the electrical current
density, $J_p=\mu_0 M_S^2|e|h/\hslash$, $h$ is the disk thickness, $e$ is the
electron charge, $\mathcal{A} =
4\eta_{sp}^{3/2}/\left[3(1+\eta_{sp})^3-16\eta_{sp}^{3/2}\right]$,
$\mathcal{B} =
(1+\eta_{sp})^3/\left[3(1+\eta_{sp})^3-16\eta_{sp}^{3/2}\right]$, and
$\eta_{sp} \in(0;1)$ denotes the degree of the spin polarization; $\sigma=\pm
1$ gives two directions of the spin-current polarization.

\section{Vortex structure and Rigid vortex dynamics}
\label{sec:vortex} %

In the case of weak dipolar interactions, the characteristic exchange length
$\ell$ is larger than the lattice constant $a$, so that in the lowest
approximation in the small parameter $a/\ell$ and weak gradients of
magnetization $\vec{m} = -\langle \vec{S}_{\vec{n}} \rangle =
\left(\sin\theta\cos\phi, \sin\theta\sin\phi,\cos\theta \right)$ we can use the
continuum approximation for the Hamiltonian \eqref{eq:Hamiltonian}.
\begin{equation} \label{eq:Energy}
\mathscr{E} = \int \mathrm{d}^2x \left[\frac{\ell^2}{2}
\left(\vec{\nabla}{\vec{m}}\right)^2 -
\frac{\vec{m}\cdot\vec{H}^{\text{ms}}}{2} - \vec{m}\cdot\vec{b}\right],
\end{equation}
where $\vec{H}^{\text{ms}}$ is the normalized magnetostatic field, which comes from
the dipolar interaction \cite{Akhiezer68}. In \eqref{eq:Energy} we suppose
that the magnetization distribution does not depend on the $z$-component (which is
valid for thin disks) and we have normalized the energy by the disk thickness $h$.

Let us consider a cylindric nanoparticle of top surface radius $L$ and
thickness $h$. For small size nanoparticles the ground state is uniform; it
depends on the particle aspect ratio $\varepsilon = h/(2L)$: thin nanodisks are
magnetized in the plane (when $\varepsilon <\varepsilon_c\approx0.906$
\cite{Aharoni90}) and thick ones along the axis (when $\varepsilon
>\varepsilon_c$). When the particle size exceeds some critical value,
typically $(3-4)\ell$ \cite{Ross02}, the magnetization curling becomes
energetically preferable due to the competition between the exchange and
dipolar interactions. For a disk-shaped particle there appears the vortex
state. For thin enough disks the vortex structure does not depend on the
$z$-coordinate of the disk and the magnetization distribution for the vortex,
which is situated in the disk center, is determined by the expressions
\cite{Mertens00}
\begin{subequations} \label{eq:vortex-ansatz-1-2}
\begin{equation} \label{eq:vortex-ansatz}
\cos\theta = pf(|\zeta|), \qquad \phi = q\arg \zeta + C \pi/2.
\end{equation}
Here $\zeta =x+{i} y$ is the coordinate in the disk plane, $q=+1$ is the
vortex $\pi_1$ topological charge (vorticity), $C=\pm1$ characterizes the
vortex chirality, and $p=\pm1$ determines the direction of the vortex core
magnetization (polarity). The bell-shaped function $f(|\zeta|)$ describes the
vortex core magnetization. The vortex polarity is connected to the $\pi_2$
topological properties of the system, the Pontryagin index
\begin{equation*} \label{eq:Ponryagin-index}
Q = \frac{1}{8\pi}\int \mathrm{d}^2x
\epsilon_{ij} \vec{m}\cdot \left[\partial_j\vec{m} \times \partial_j
\vec{m} \right].
\end{equation*}
For the vortex configuration the Pontryagin index takes the half-integer
values $Q=-pq/2$.

For the vortex solution localized at $Z=X+{i} Y=R\exp({i}\Phi)$, the
$\phi$-field has the form $\phi = q\arg (\zeta-Z) + C \pi/2.$ Such a form of
the $\phi$--field satisfies the Laplace equation, so it describes the planar
vortex dynamics (neglecting the z-component of magnetization) for the infinite
Heisenberg magnets. The finite size effects for the circular Heisenberg magnet
can be described by the image-vortex ansatz \cite{Mertens00}. Moreover, one has
to take into account the long-range dipolar interactions, which lead in
general to integro-differential equations \cite{Guslienko04}. For a thin
ferromagnet the dipolar interactions produce an effective uniaxial anisotropy
of the easy-plane type caused by the faces surface magnetostatic charges
\cite{Gioia97,Ivanov02a} and an effective in-plane anisotropy caused by the
edge surface charges (surface anisotropy) \cite{Carbou01,Ivanov02a}. Due to
the surface anisotropy the magnetization near the disk edge is constrained to
be tangential to the boundary, which prevents its precession near the edge
(absence of surface charges) \cite{Ivanov02a,Guslienko05,Caputo07b}. Finally, the
$\phi$--field can be written in the form
\begin{equation} \label{eq:phi-ansatz}
\phi= \arg\bigl[\zeta-Z\bigr] + \arg\bigl[\zeta-Z^I\bigr] - \arg Z +C \frac{\pi}{2}.
\end{equation}
\end{subequations}
Here $Z^I = Z L^2/R^2$ is the image vortex coordinate; the image vortex is added to satisfy the Dirichlet boundary conditions.

Under the influence of an external force the vortex starts to move. If
such forces are weak, the vortex behaves like a particle during its
evolution. Such rigid vortex dynamics can be well-described using the
Thiele approach \cite{Thiele73,Huber82}. Following this approach we
use the traveling wave ansatz $\vec{m}(\vec{r},\tau) =
\vec{m}(\vec{r}-\vec{R}(\tau))$, where the function $\vec{m}$ on the
right--hand side describes the vortex shape \eqref{eq:vortex-ansatz-1-2}.
This results in a force balance equation $\vec{F}^g + \vec{F}^d + \vec{F} = 0$.
The first term $\vec{F}^g = -2Q\left[\dot{\vec{R}} \times
\hat{\vec{z}}\right]$ is a gyroscopical force
\cite{Huber82,Nikiforov83}, which acts on a moving vortex in the
same way as the magnetic field influences a charged particle by a
Lorentz force. The second term is a dissipative force
$\vec{F}^d=-\eta \dot{\vec{R}}$ with $\eta\approx(\alpha/2)\ln L$,
and the last term is an external force
$\vec{F}=-(2\pi)^{-1}\vec{\nabla}_{\vec{R}}\mathscr{E} =
\vec{F}^{\text{ms}} + \vec{F}^{\text{ext}}$, where the total energy is
described by Eq.~\eqref{eq:Energy}. The magnetostatic part of this
force is caused mainly by magnetostatic energy of volume charges.
For small displacements $R\ll L$ from the disk origin and small
aspect ratios ($\varepsilon \ll1$), it can be found analytically that $\vec{F}^{\text{ms}} =
-\Omega_G \vec{R}$ \cite{Guslienko02a} with $\Omega_G\sim 10\varepsilon/9 \pi$
\cite{Guslienko06}. For finite $\varepsilon$ one can use the
analysis in Ref.~\cite{Ivanov04b,Zaspel05}, which results in
$\Omega_G(\varepsilon)$. Finally the force balance condition takes the form of a Thiele equation
\begin{equation} \label{eq:force-balance}
2Q\left[\dot{\vec{R}} \times \hat{\vec{z}}\right] = \vec{F}^{\text{ext}}
- \eta \dot{\vec{R}} - \Omega_G \vec{R}.
\end{equation}
Without dissipation and external force this equation describes a gyration
of the vortex around the disk origin with a frequency $\Omega_G$.
The dissipation damps the possible circular motion of the vortex. However, an
external force can excite a non-decaying vortex motion. For example, it is
known that the rotating magnetic field can excite a circular rotation of the
vortex in a Heisenberg magnet \cite{Zagorodny04,Sheka05} and a nanomagnet
\cite{Kravchuk07c} due to the force \cite{Sheka05}
\begin{subequations} \label{eq:force}
\begin{equation} \label{eq:force-field}
\vec{F}^{\text{ext}}\! = \!
-\frac{bL}{2R}\Bigl\{\!\vec{R}\cos(\Phi-\omega\tau) +
\left[\vec{R} \times
\hat{\vec{z}}\right]\sin(\Phi-\omega\tau)\!\Bigr\}.
\end{equation}
In the case of electrical current influence the
circular vortex dynamics can be excited by the force
\cite{Liu07b,Ivanov07b,Sheka08a}:
\begin{equation} \label{eq:force-sc}
\vec{F}^{\text{ext}} = \frac{j\sigma\mathcal{A}q}{2\omega_0}
\left[\vec{R} \times \hat{\vec{z}}\right].
\end{equation}
\end{subequations}
Note that the vortex motion can be excited by any small magnetic field
$(b,\omega)$ \cite{Kravchuk07c}; howeber, to excite the vortex
motion by a spin current, its intensity should exceed threshold value \cite{Ivanov07b,Sheka08a}.

\section{Numerical studies}
\label{sec:numerics} %

In order to check the analytical predictions about the driven vortex dynamics,
we have performed numerical simulations. In the case of a rotating magnetic field we
used the micromagnetic simulator for Landau-Lifshitz-Gilbert equations
\cite{OOMMF} as described in Ref.~\cite{Kravchuk07c}. To study the vortex
dynamics under the action of a spin-polarized electrical current, we used
numerical simulations of the discrete spin-lattice Eq.~\eqref{eq:LLS-discrete}
with the spin torque given by Eq.~\eqref{eq:T} as described in
Refs.~\cite{Sheka07b,Sheka08a}. In both cases, as initial condition we use
the vortex with positive polarity ($p=+1$), centered at the disk origin.
To identify precisely the vortex position we use, similar to
Ref.~\cite{Hertel06}, the crossection of isosurfaces $m_x=0$ and $m_y=0$.

Let us start with the case of a spin current. If we apply the current, whose
polarization is parallel to the vortex polarity ($j\sigma p>0$), the vortex
does not quit the disk center, which is a stable point. However, if the spin
current has the opposite direction of the spin polarization ($j\sigma p<0$,
$p=+1$, $\sigma=+1$ and $j<0$ in our case) the vortex motion can be excited
when the current intensity is above a threshold value
\cite{Liu07b,Ivanov07b,Sheka08a} as a result of the balance between the
pumping and damping. Following the spiral trajectory, the vortex finally
reaches a circular limit cycle. The rotation sense of the spiral is determined by
the gyroforce, i.e. by the topological charge $Q$, which is a clockwise rotation in our case. At some value, the radius becomes comparable with the system size
$L$ and the vortex dynamics becomes more complicated and can not be described by the Thiele equation \eqref{eq:force-balance}. Finally it results in a
switching of the vortex core, see below.

A qualitatively similar picture is found in the case of a rotating field
with angular frequency directed opposite to the vortex polarity: $\omega p
<0$. There exist different regimes in the vortex dynamics
\cite{Kravchuk07c}. In the range of frequencies close to the
frequency of the orbital vortex motion ($\Omega\sim 1$ GHz) the
vortex demonstrates a finite motion in a region near the disk center
along a quite complicated trajectory and does not switch its
polarity. These cycloidal vortex oscillations are similar to those in Ref.~\cite{Kovalev03a}, and correspond to the excitation of higher magnon
modes during the motion. If $\omega\gg\Omega$, then for weak fields
the vortex motion can be considered as a sum of two constituents:
(i) the gyroscopic orbital motion as without field, and (ii)
cycloidal oscillations caused by the field influence. For the case
$\omega p<0$ the direction of the cycloidal oscillations coincides with the direction of the
field rotation, while the direction of the gyroscopic orbital motion is opposite to it.
For stronger fields the vortex motion becomes more complicated
and its average motion can be directed even opposite to the gyroscopical motion
(this situation is shown in Fig.~\ref{fig:V_AV_NV}).
The irreversible switching of the vortex polarity can be excited in a specific range of parameters $(\omega,b)$, with typical frequencies about $10$ GHz and intensities about $20$ mT \cite{Kravchuk07c}. The mechanism of the switching is discussed below.

\section{Vortex polarity switching}
\label{sec:switching}

\begin{figure*}
\includegraphics[width=0.25\textwidth]{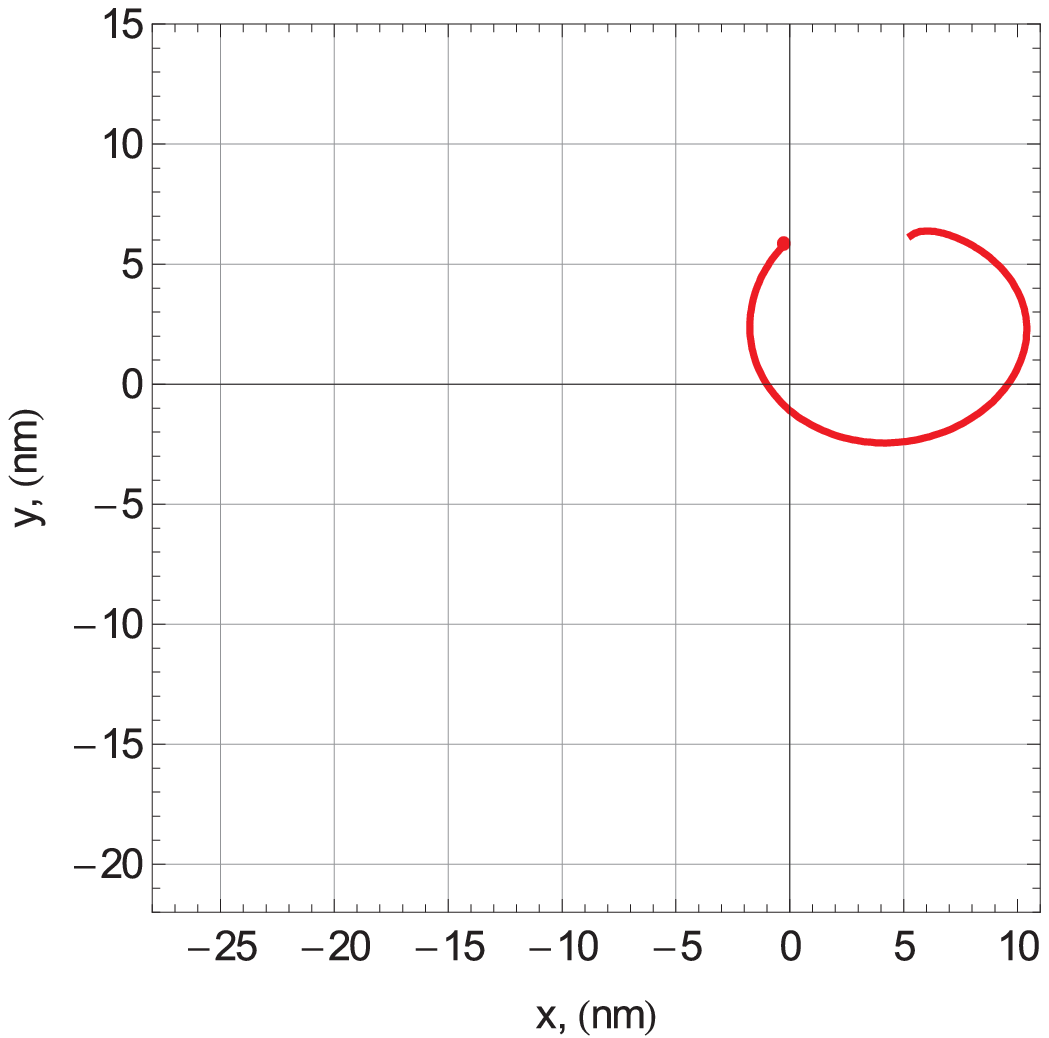}
\includegraphics[width=0.25\textwidth]{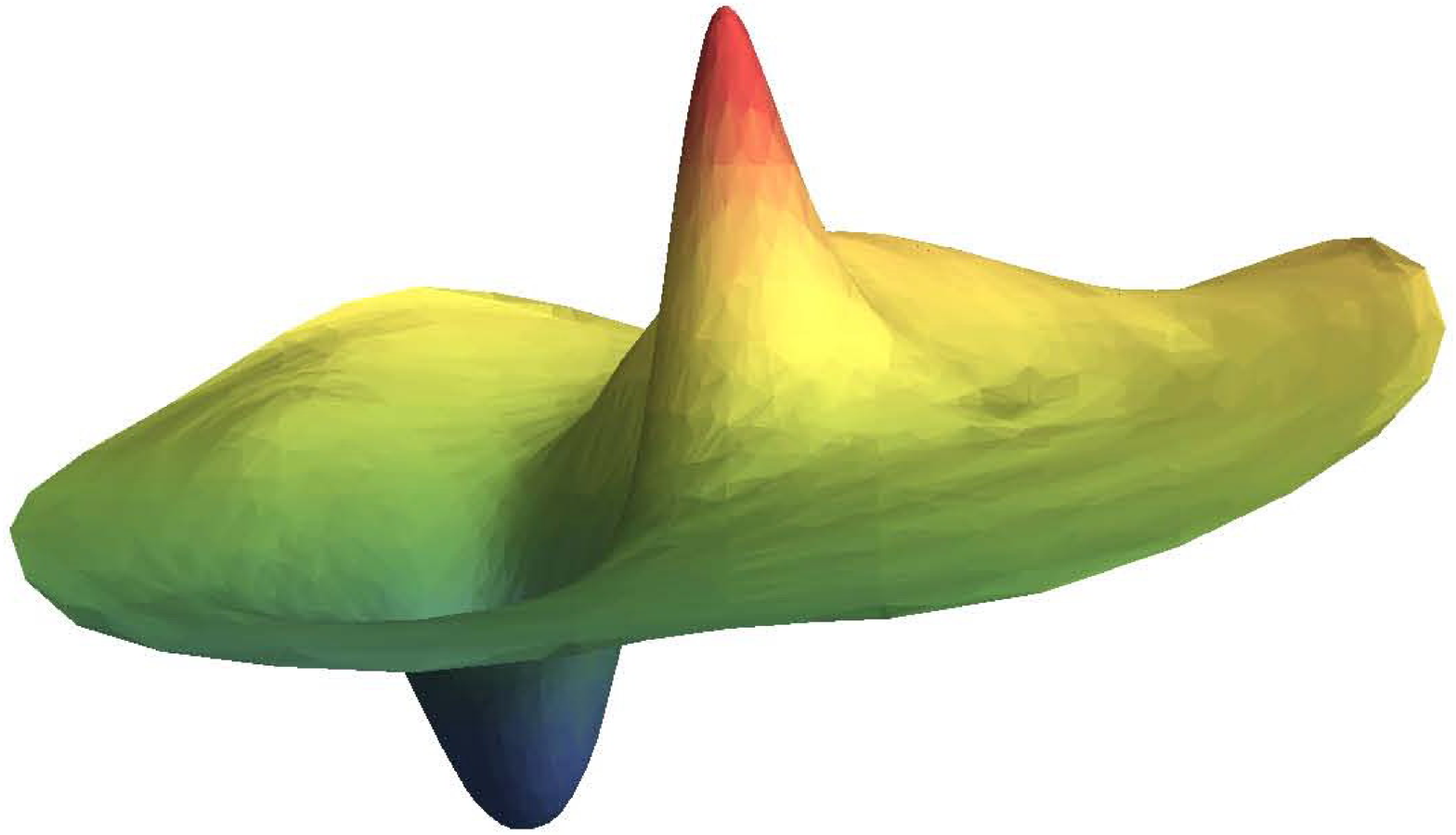}
\includegraphics[width=0.25\textwidth]{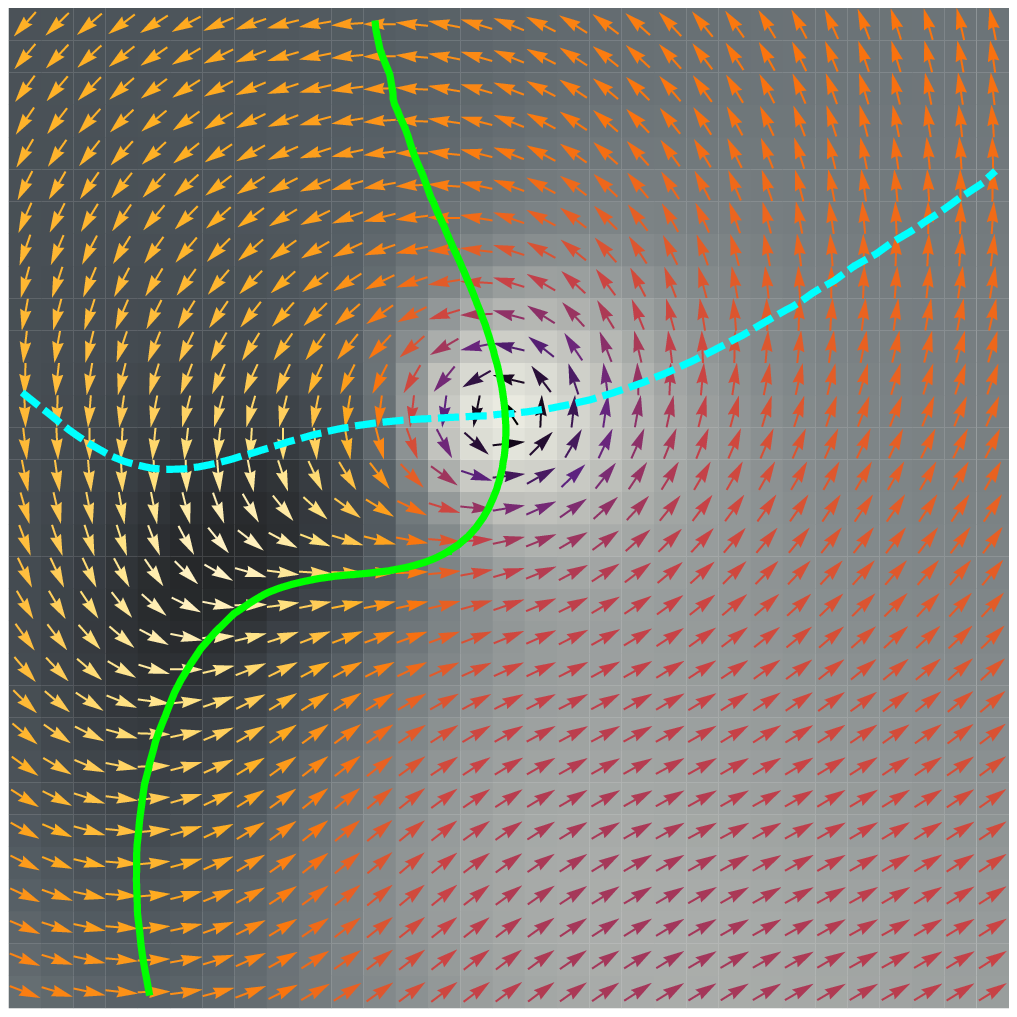}

(a)~\textbf{t=110 ps}: the rotating field causes the vortex motion, as a result a negative dip is forming.

\includegraphics[width=0.25\textwidth]{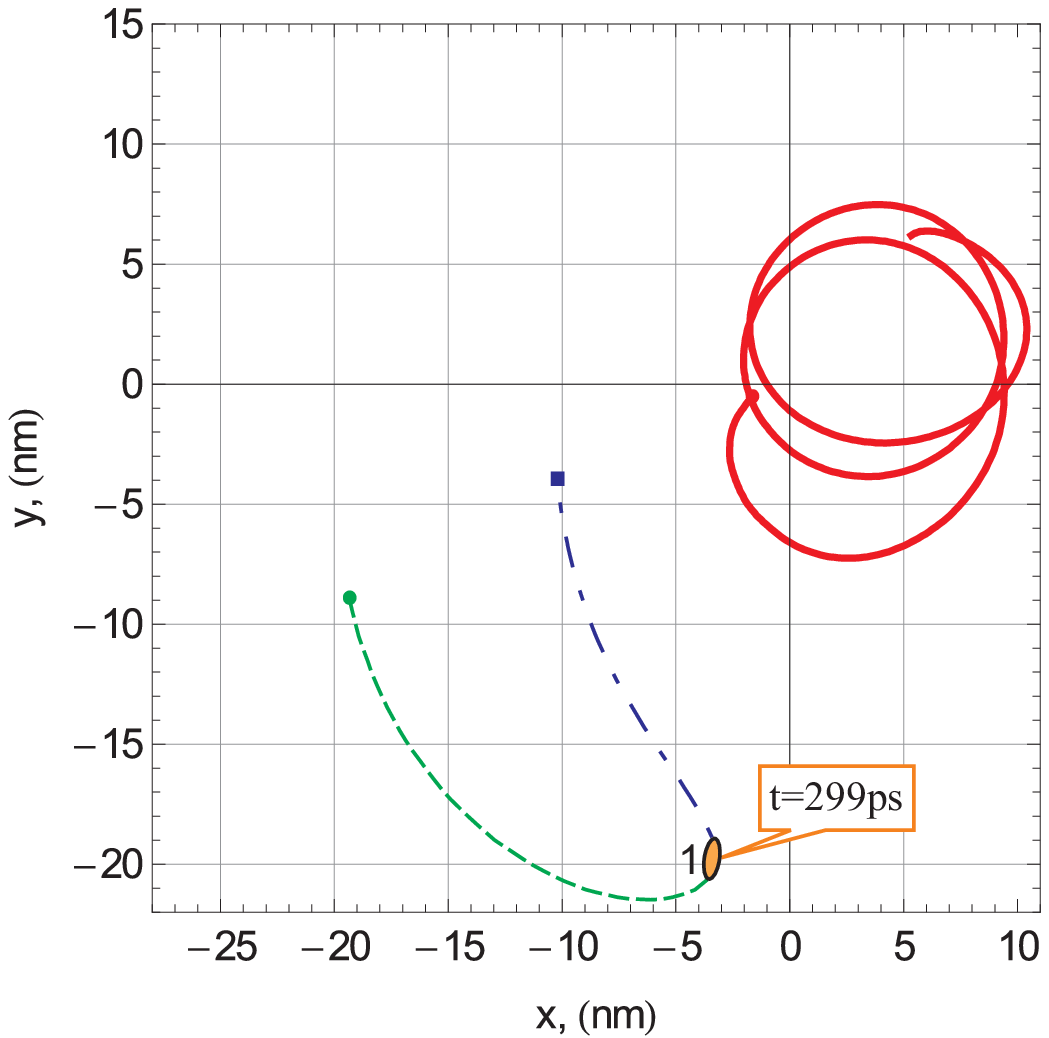}
\includegraphics[width=0.25\textwidth]{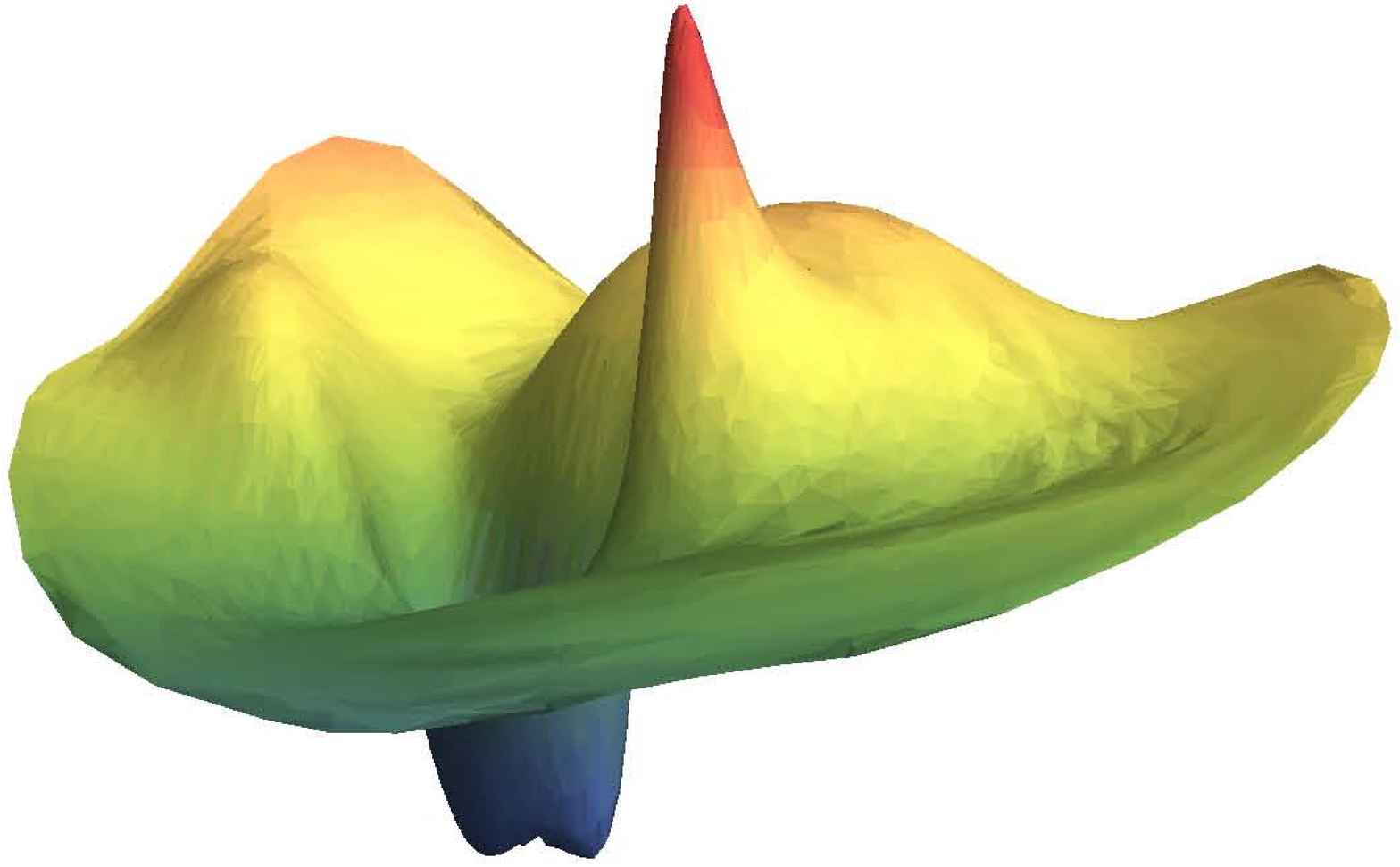}
\includegraphics[width=0.25\textwidth]{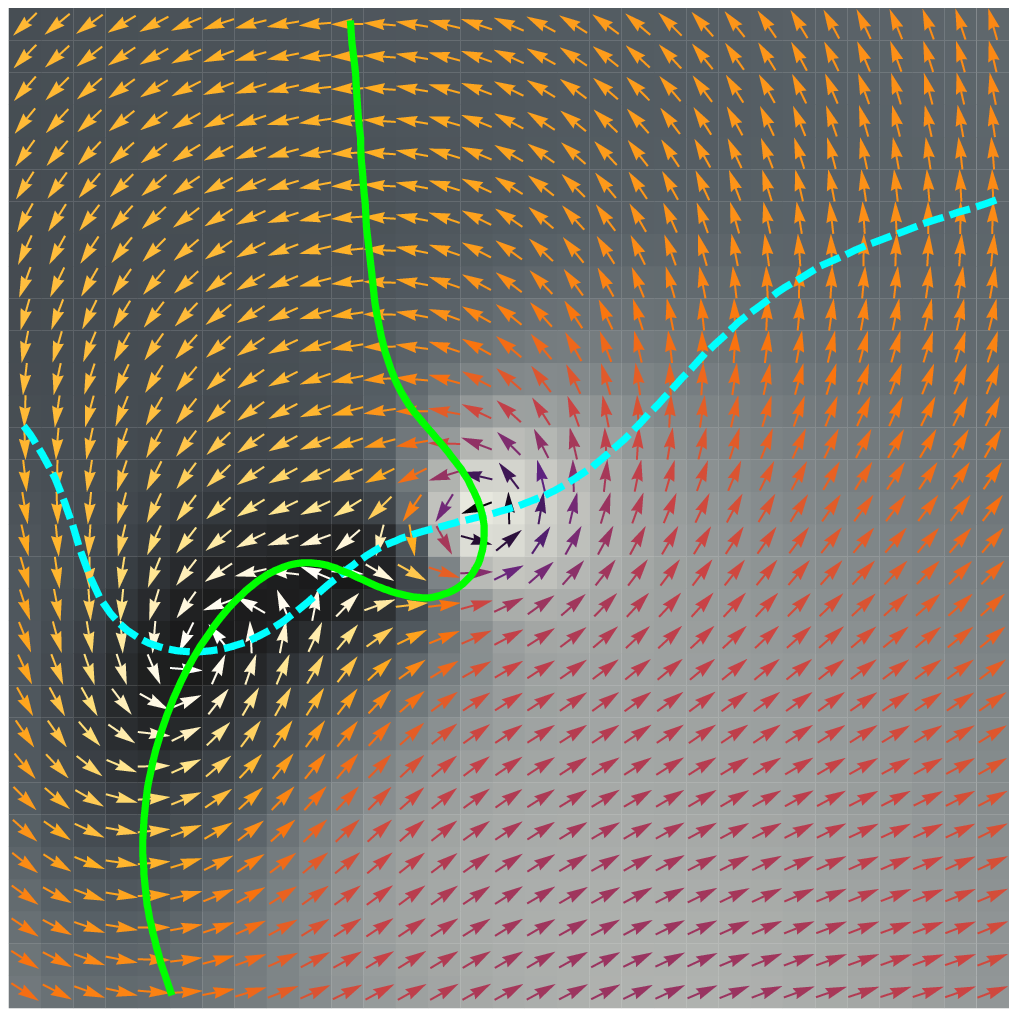}

(b)~\textbf{t=320 ps}: a vortex-antivortex pair has been born in region 1.

\includegraphics[width=0.25\textwidth]{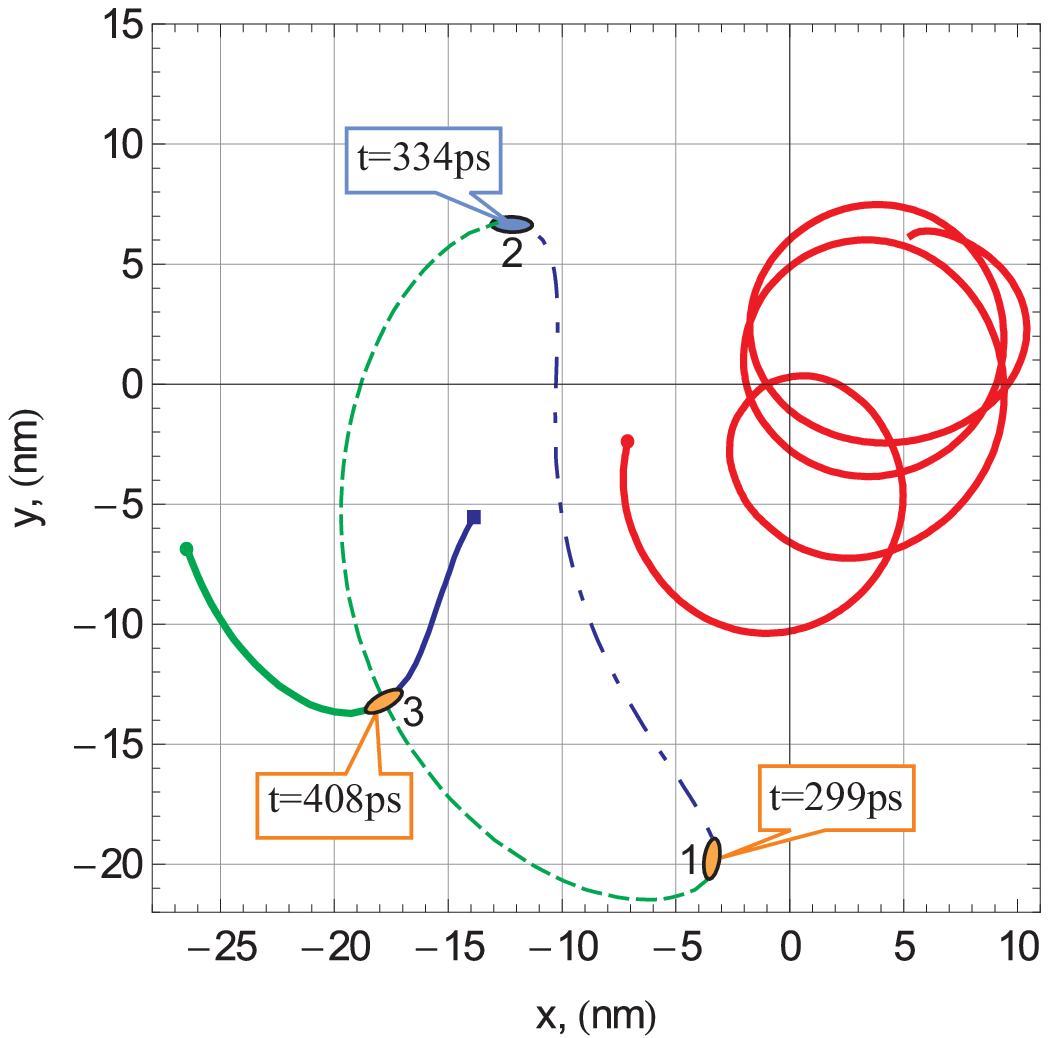}
\includegraphics[width=0.25\textwidth]{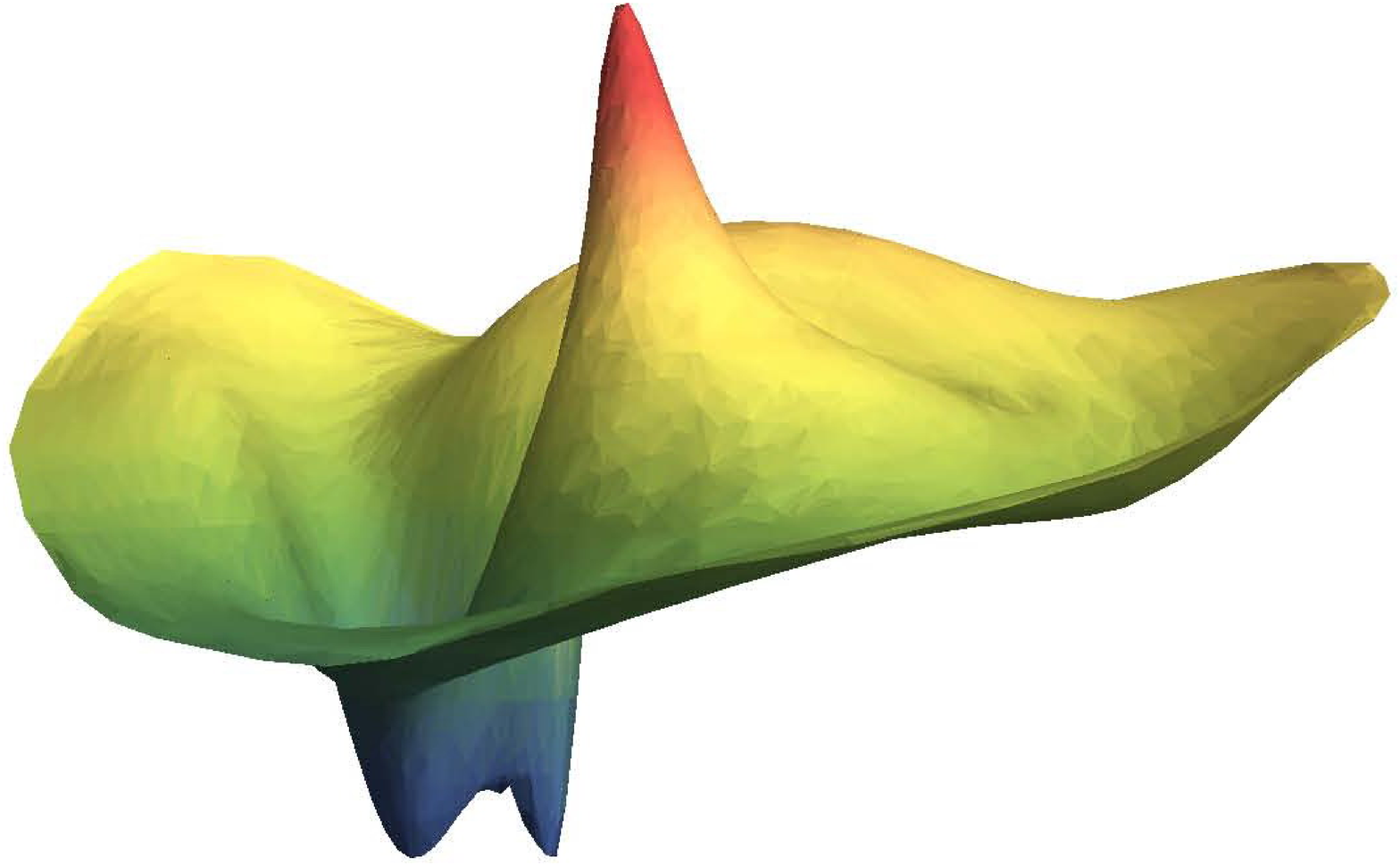}
\includegraphics[width=0.25\textwidth]{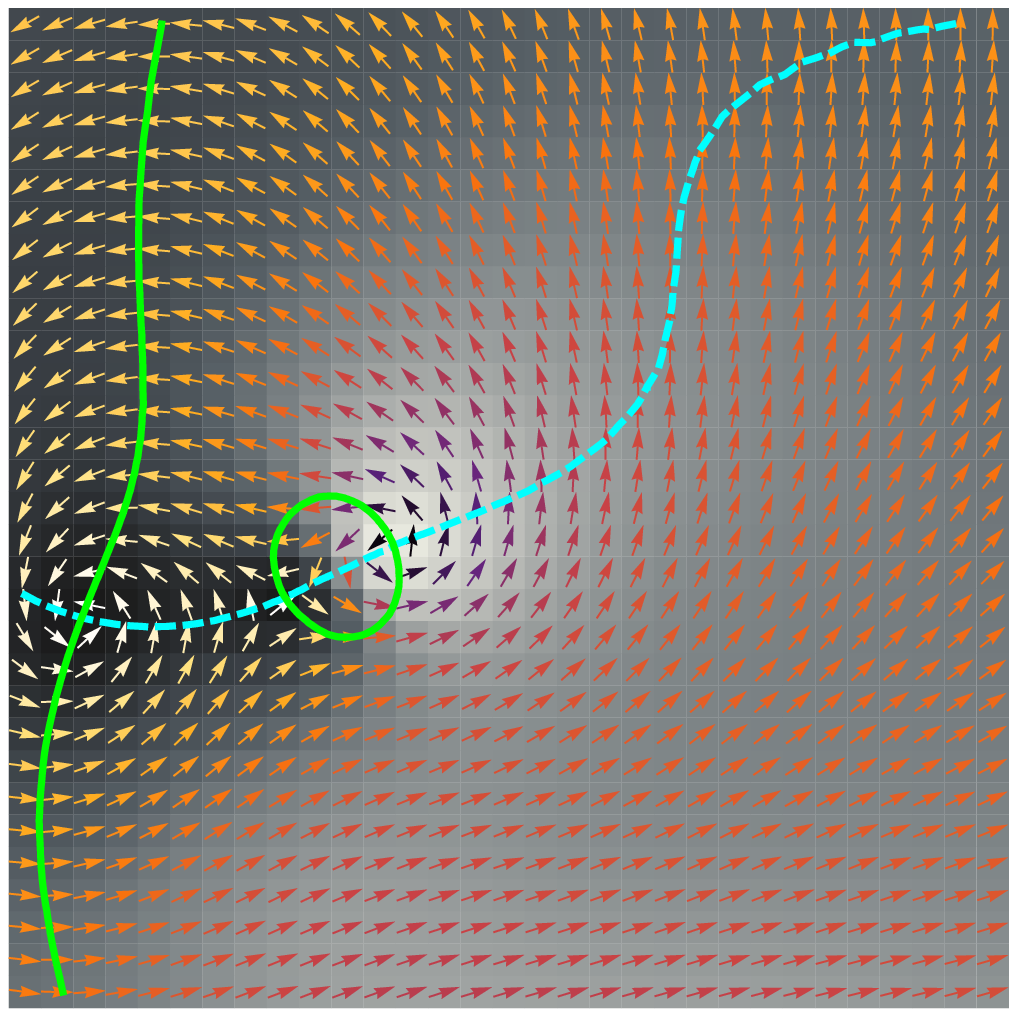}

(c)~\textbf{t=420 ps}: the vortex pair has annihilated in
region 2, after some time a new vortex-antivortex pair is born in
region 3.

\includegraphics[width=0.25\textwidth]{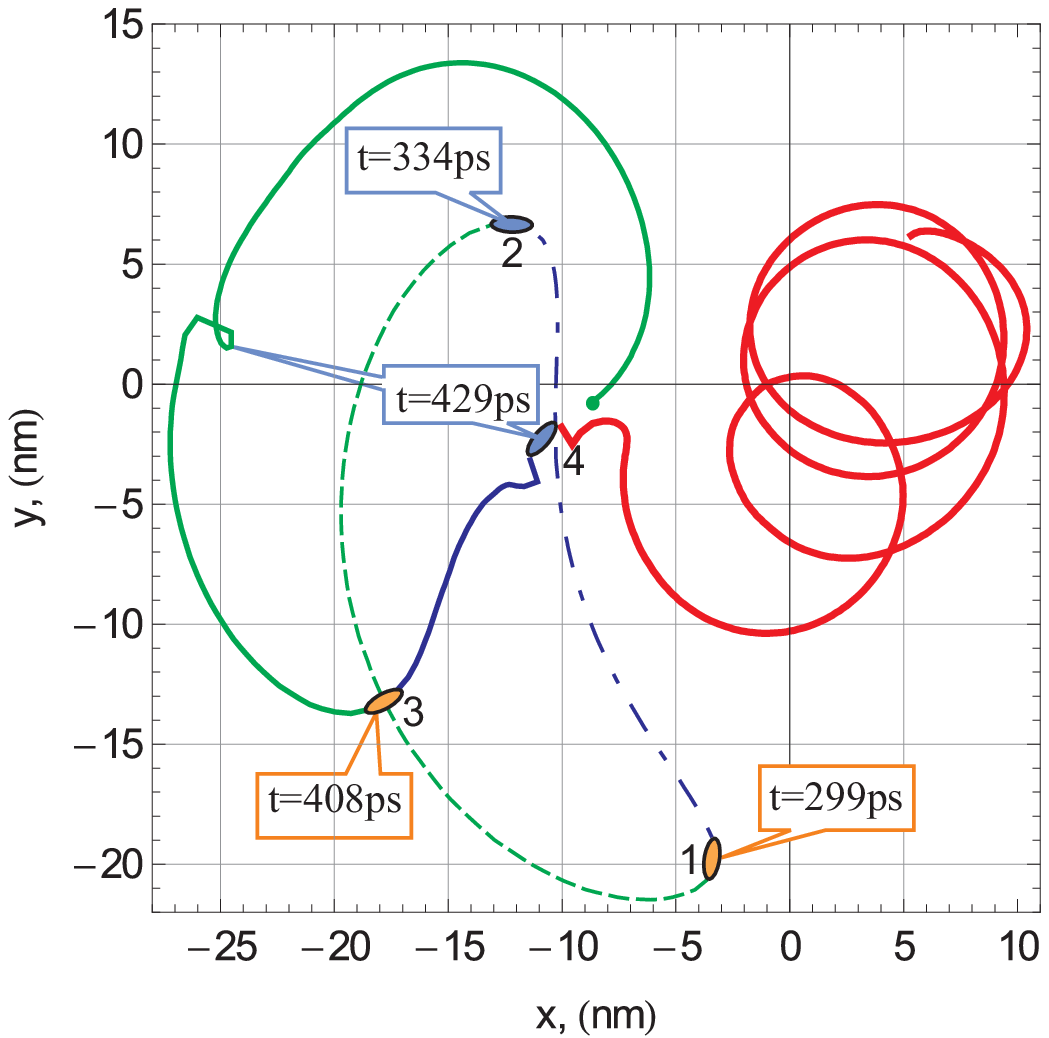}
\includegraphics[width=0.25\textwidth]{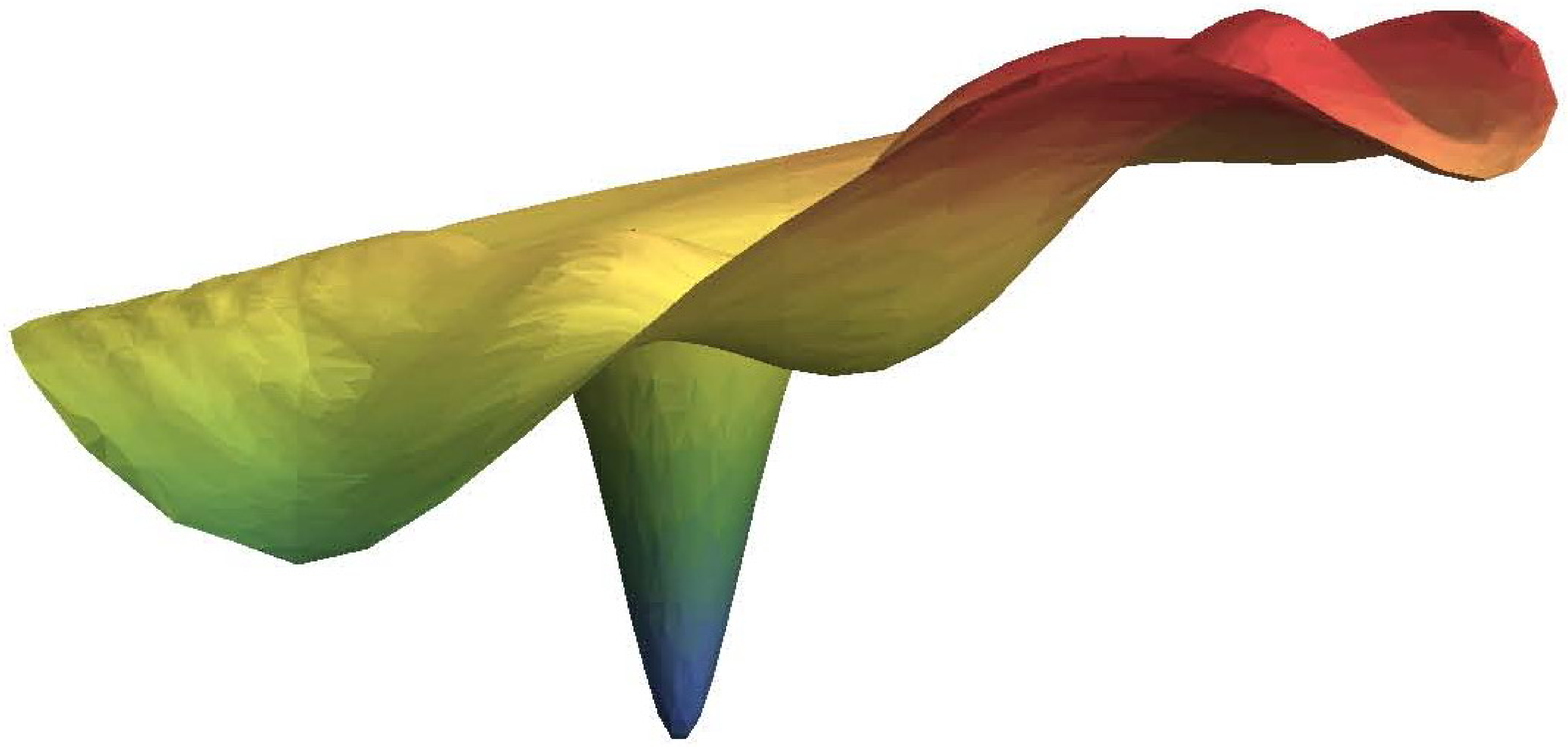}
\includegraphics[width=0.25\textwidth]{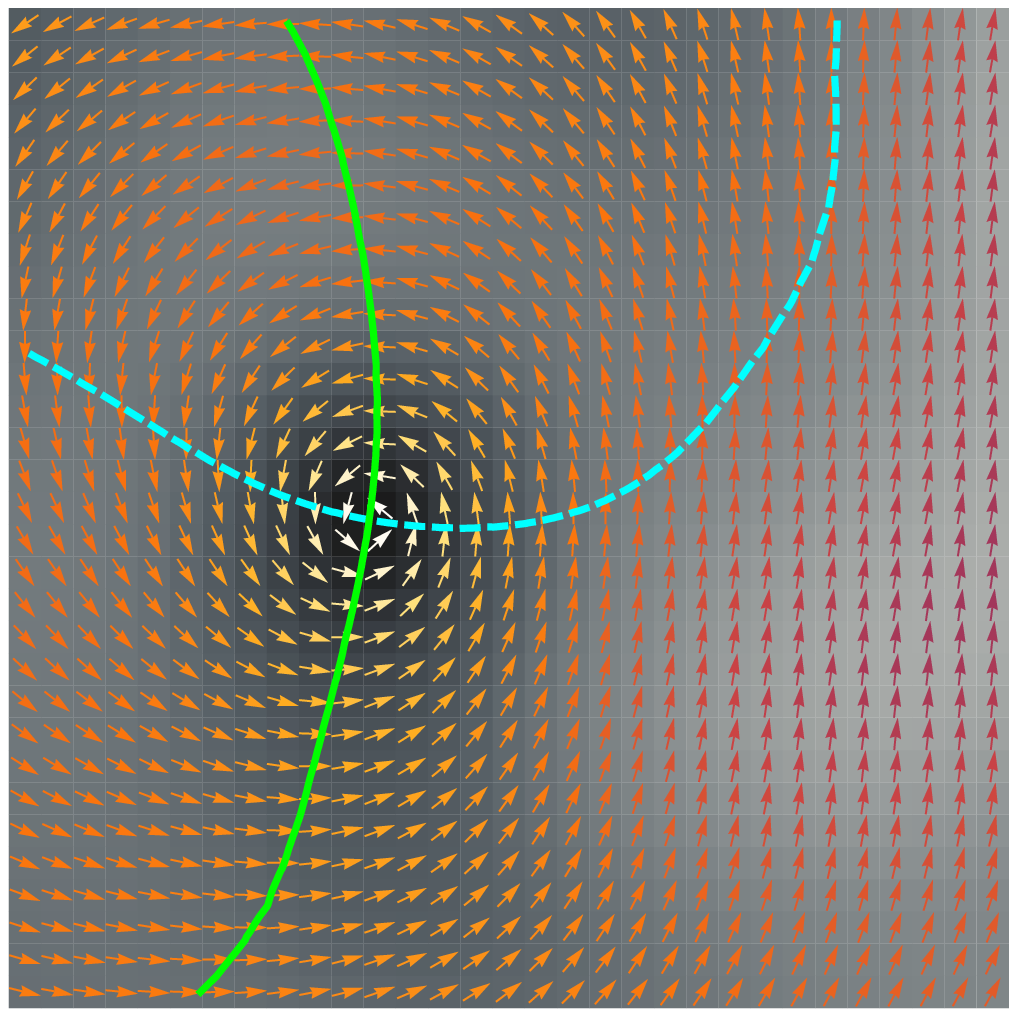}

(d)~\textbf{t=495 ps}: the new antivortex has annihilated with the old vortex in region 4 and only the new vortex of opposite polarity remains.

\caption{The vortex switching process under rotating field influence. Left column -- vortex trajectories. Red line -- initial vortex core trajectory, green line -- new vortex, blue line -- antivortex. Right column -- in-plane magnetization distribution, the out-of-plane component is denoted by levels of gray color, dashed blue and solid green lines correspond to isosurfaces $m_x=0$ and $m_y=0$, respectively. The out-of-plane component $m_z$ is shown in the middle column. The data were obtained from micromagnetic simulation for a permalloy disk (132nm diam., 20nm thickness) dynamics under rotating field ($B=0.07$~T, $\omega=10$~GHz) influence.}

\label{fig:V_AV_NV}%
\end{figure*}

The mechanism of the vortex switching is of general nature; it is
essentially the same in all systems where the switching was observed
\cite{Waeyenberge06,Xiao06,Hertel07,Lee07a,Yamada07,Kravchuk07c,Sheka07b,Liu07b}.
Under the action of pumping, the original vortex (V,
$q_V=1,p_V=1,Q_V=-1/2$), situated in the disk center,
moves along a spiral trajectory in the case of a current or along a
more complicated trajectory (as described in the previous section)
in the case of the rotating field. During its motion, the
vortex excites a number of spin waves \cite{Choi07}. Mainly magnon
modes of two kinds are excited in this system: symmetrical and
azimuthal ones \cite{Gaididei00,Zagorodny03}. Due to the continuous
pumping the system goes to the nonlinear regime: the amplitude of
the azimuthal mode increases and there appears an out-of-plane dip
nearby the vortex, see Fig.~\ref{fig:V_AV_NV}(a). When the amplitude
of the out-of-plane dip reaches its minimum [$m_z=-1$,
Fig.~\ref{fig:V_AV_NV}(b)] a pair of a \emph{new} vortex (NV,
$q_{NV}=-1,p_{NV}=-1,Q_{NV}=-1/2$) and antivortex (AV,
$q_{AV}=1,p_{AV}=-1,Q_{AV}=1/2$) is created. These three objects move
following complicated trajectories which result from Thiele-like
equations. The directions of the motion of the partners are
determined by the competition between the gyroscopical motion and
the external forces, which are magnetostatic force $F^{\text{ms}}$, pumping
force $F^{\text{ext}}$, see Eqs.~\eqref{eq:force}, and the interactions between the
vortices $\vec{F}_i^{\text{int}}$.

The magnetostatic force for the three body system can be calculated, using the
image vortex approach, which corresponds to fixed (Dirichlet) boundary conditions. For the vortex state nanodisk such
boundary conditions results from the magnetostatic interaction,
which is localized near the disk edge \cite{Ivanov02a}. The same
statement is also valid for the three vortex (V-AV-NV) state, which
is confirmed also by our numerical simulations. The $\phi$--field can be presented by the three vortex ansatz
\begin{equation} \label{eq:three-phi-ansatz}
\phi= \sum_{i=1}^3 q_i \Bigl\{ \arg\bigl[\zeta-Z_i\bigr]
+ \arg\bigl[\zeta-Z_i^I\bigr] - \arg Z_i\Bigr\} +C \frac{\pi}{2}.
\end{equation}
The force coming from the volume magnetostatic charge density $\lambda = -\vec{\nabla}\cdot \vec{m}$ can be calculated in the same way as for a single vortex \cite{Metlov02,Ivanov07b}, which results in
$\vec F^{\text{ms}}\approx-\Omega_G q_i\sum_jq_j\vec{R_j}$, where $q_i=\pm1$ is the vorticity of $i$-th vortex (antivortex).

The interaction force $\vec{F}^{\text{int}}$ between vortices is a 2D coulomb force $\vec{F}_i^{\text{int}} = \sum\limits_{i\neq j}q_i q_j \frac{\vec{R}_i -
\vec{R}_j}{\left|\vec{R}_i - \vec{R}_j\right|^2}$.

The internal gyroforces of the dip impart the initial velocities to the NV and AV, which
are perpendicular to the initial dip velocity and have opposite
directions. In this way the AV gains the velocity component directed
to the center, where the V is moving. This new-born pair is a topologically trivial
$Q=0$ pair, which undergoes a Kelvin motion \cite{Papanicolaou99}. This Kelvin pair collides with the original vortex. If the pair was born far from
V then the scattering process is semi--elastic; the new-born pair is not destroyed by the collision. In the exchange approach this pair survives, but it is scattered by some angle \cite{Komineas07b}. Due to additional forces (pumping, damping and magnetostatic interaction), the real motion is more complicated and the Kevin pair can finally annihilate by itself, see Fig.~\ref{fig:V_AV_NV}(c).

Another collision mechanism takes place when the pair is born closer to the original vortex. Then the AV can be captured by the V and an annihilation with the original vortex happens, see Fig.~\ref{fig:V_AV_NV}(d). The collision process is essentially inelastic. During the collision, the original vortex and the antivortex form a topological nontrivial
pair ($Q=-1$), which performs a rotational motion around some guiding
center \cite{Komineas07a,Komineas07b}. This rotating vortex dipole forms a
localized skyrmion (Belavin-Polyakov soliton \cite{Belavin75}),
which is stable in the continuum system. In the discrete lattice
system the radius of this soliton, i.e. the distance between vortex and antivortex, rapidly decreases almost without energy loss. When the soliton radius is about one lattice constant, the pair undergoes the topologically forbidden annihilation \cite{Kravchuk07c,Komineas07a}, which is accompanied by strong spin-wave radiation, because the topological properties of the system change \cite{Hertel06,Tretiakov07}.

The three--body problem can be analytically described in a rigid vortex approach, based on Thiele-like equations \eqref{eq:force-balance}:
\begin{equation} \label{eq:force-balance-2}
\begin{split}
&2Q_i\left[\dot{\vec{R}}_i \times \hat{\vec{z}}\right] = \sum_{i\neq
j}q_i q_j \frac{\vec{R}_i - \vec{R}_j}{\left\lvert
\vec{R}_i - \vec{R}_j \right\rvert^2} + \vec{F}_i^{\text{ext}} \\
& - \eta \dot{\vec{R}}_i - \Omega_G(\varepsilon)q_i\sum\limits_j
q_j\vec R_j, \qquad i=1,2,3.
\end{split}
\end{equation}
Here $ \vec{F}_i^{\text{ext}}$ is the driving force \eqref{eq:force}, caused by the field \eqref{eq:force-field} or spin-current \eqref{eq:force-sc} pumping. This three--body process in exchange interaction approximation is studied in detail by \citet{Komineas07b}.

\begin{figure}
\includegraphics[width=\columnwidth]{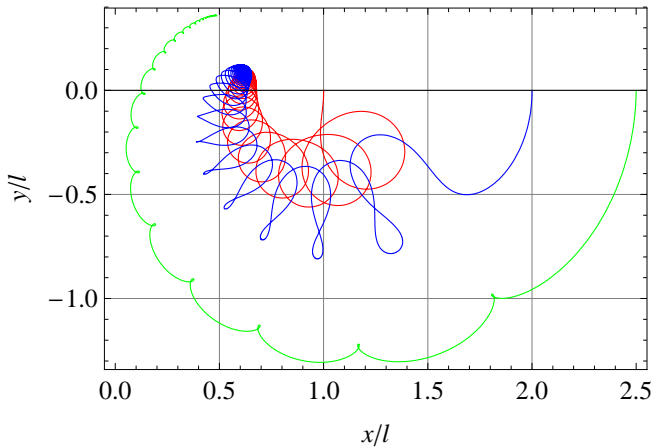}
\caption{Three--body dynamics from the solution of Eq.~\eqref{eq:force-balance-2}, for a system with $L/\ell = 20$, $\alpha = 0.01$, $b=0.02$, $\omega=1$, and $\Omega_G = 0.0287$. The original vortex (red line) was positioned at $(1,0)\ell$, the antivortex (blue) at $(2,0)\ell$, and the newborn vortex (green) at $(2.5, 0)\ell$. }
\label{fig:R_V_AV}%
\end{figure}

The set of Eqs.~\eqref{eq:force-balance-2} describes the main features of the observed three--body dynamics. During the evolution, the original vortex and the antivortex create a rotating dipole, in agreement with Refs.~\cite{Komineas07a,Komineas07b}, see Fig.~\ref{fig:R_V_AV}. The distance in this vortex dipole rapidly tends to zero. The new--born vortex moves on a clockwise spiral to the origin.

\section{Summary}

To summarize, we have studied the switching of the magnetic vortex polarity under the action of a rotating magnetic field and under the action of a spin-polarized current. The switching picture involving the creation and annihilation of a vortex-antivortex pair is very general and does not depend on the details how the vortex dynamics was excited. In particular, such a switching mechanism can be induced by a field pulse \cite{Waeyenberge06,Xiao06,Hertel07}, by an AC oscillating \cite{Lee07a} or rotating field \cite{Kravchuk07c}, by an in-plane electrical current (nonhomogeneous spin torque) \cite{Yamada07,Liu07a} and by a perpendicular current (homogeneous spin torque) \cite{Sheka07b,Liu07b,Ivanov07b}. Our analytical analysis is confirmed by numerical spin-lattice simulations.

The authors acknowledge support from Deutsches Zentrum f{\"u}r Luft- und
Raumfart e.V., Internationales B{\"u}ro des Bundesministeriums f{\"u}r
Forschung und Technologie, Bonn, in the frame of a bilateral scientific
cooperation between Ukraine and Germany, project No.~UKR~05/055. Yu.G., V.P.K.
and D.D.S. thank the University of Bayreuth, where a part of this work was
performed, for kind hospitality. Yu. G. acknowledges support from the Special
Program of Department of Physics and Astronomy of the National Academy of
Sciences of Ukraine. V.P.K. and D.D.S. acknowledge support from the grant
No.~F25.2/081 from the Fundamental Researches State Fund of Ukraine. D.D.S.
acknowledges also the support from the Alexander von Humboldt--Foundation.


\begin{thebibliography}{72}
\bibitem[{Kov()}]{Kovalev-vortex}
\bibinfo{note}{To the best of our knowledge the term \emph{magnetic vortex} was
  introduced by the Kosevich group in Ref.~\cite{Kovalev79} for the object,
  which is now called \emph{magnetic skyrmion} and is widely used for the
  localized solitons, e.g. in easy-axis magnets.}
\end{thebibliography}

\end{document}